\documentclass[aps,prb,twocolumn,superscriptaddress,amsmath]{revtex4} 
\usepackage{makeidx}

\usepackage[colorlinks,
hyperindex,
colorlinks,%
citecolor=blue,%
linkcolor=blue,%
urlcolor=blue, %
]{hyperref}
\usepackage[dvipsnames]{xcolor}
\usepackage[pdftex]{graphicx}  
\usepackage{subfigure}
\usepackage{epstopdf}
\usepackage{dcolumn}   
\usepackage{bm}        
\usepackage{amssymb}   
\usepackage{amsmath}
\usepackage{newtxtext,newtxmath}
\usepackage{siunitx}

\usepackage{textcomp}

\makeindex

\begin{document}
	
	

	\title{Phonon anharmonic frequency shift induced by four-phonon scattering calculated from first principles}
	
	\author{Tianli Feng}
	\email{tianli.feng01@gmail.com}
	\affiliation{School of Mechanical Engineering and the Birck Nanotechnology Center, Purdue University, West Lafayette, Indiana 47907-2088, USA}
	
	\author{Xiaolong Yang}
	\affiliation{School of Mechanical Engineering and the Birck Nanotechnology Center, Purdue University, West Lafayette, Indiana 47907-2088, USA}
	\affiliation{Frontier Institute of Science and Technology, and State Key Laboratory for Mechanical Behavior of Materials, Xi'an Jiaotong University, Xi'an 710049, P. R. China.}
	
	\author{Xiulin Ruan}
	\email{ruan@purdue.edu}
	\affiliation{School of Mechanical Engineering and the Birck Nanotechnology Center, Purdue University, West Lafayette, Indiana 47907-2088, USA}
	\date{\today}
	

\begin{abstract}


Phonon energies at finite temperatures shift away from their harmonic values due to anharmonicity. In this paper, we have realized the rigorous calculation of phonon energy shifts of silicon by three and four-phonon scattering from first principles. The anharmonic fourth-order force constants are calculated by considering up to the fifth nearest neighbors. The results agree reasonably well with available data from inelastic neutron scattering throughout the Brillouin zone. Surprisingly, the frequency shifts of optical phonon modes near the $\Gamma$ point are sensitive to the cutoff radius of the fourth-order force constants, in contrast to the four-phonon scattering rates, which nearly saturate when considering the second nearest neighbors. We have also compared the results with ab initio molecular dynamics simulations and found that the higher order of anharmonicity is important for optical phonons. Our work provides critical insight into the anharmonic phonon frequency shift and will have significant impact on the thermal and optical applications.



\end{abstract}
\maketitle

\section{Introduction}


In general, phonon frequencies of solids at finite temperatures shift away from their harmonic values due to anharmonicity\,\cite{Maradudin1962pr}. Accurate prediction of the frequency shift is crucial for the study of the thermal, thermodynamical and acoustic properties as phonons carry most of heat and are responsible for most of the entropy in semiconductors and insulators in general cases\,\cite{Wallace1998,Maradudin1961,Fultz2010247}. Since the phonon frequency affects the phonon interaction with other fundamental particles, the study of frequency shift is also important in other areas such as superconductivity\,\cite{Liu2011PRL,Yildirim2001prl,Choi2002prb,Bonini2007}, infrared spectroscopy\,\cite{Balkanski1983,Bao2012jqsrt}, gravitational wave detector\,\cite{Park1997}, etc.


Anharmonicity softens phonon modes via anharmonic three and four phonon scatterings. 
In the past a few decades, several works have performed the density functional theory calculations of the frequency shift based on perturbation theory and obtained reasonable results as compared to the experimental Raman spectroscopy measurement\,\cite{Maradudin1962pr, Park1997,Lang1999,Shukla2003prl,PRB2003MgB2,Bonini2007,Nikolenko2013}. Their calculations, however, were limited at the $\Gamma$ point, which is not responsible for the thermal properties such as the thermal conductivity. Recently, Turney \textit{et al.}\,\cite{Turney2009prb1,Turney2009thesis} conducted the calculations at other points in the Brillouin zone (BZ), however, it was based on a classical interatomic potential. Another method that can extract the frequency shift from ab initio molecular dynamics is the spectral energy density analysis\,\cite{Koker2009prl,Thomas2010prb,Feng2014Jn,Feng2015jap}, which is however, quite time-consuming. Moreover, these MD-based methods cannot conveniently separate effects of three and four-phonon scatterings. 
The goal of this work is to calculate the phonon frequency shift in the full Brillouin zone based on perturbation theory using first principles. Silicon is taken as the benchmark material since it has experimental data available, which are very limited for other materials. Also, the thermal properties of silicon at finite temperature are of great importance in electronics\,\cite{green2001efficient,cui2003high}, photovoltatics\,\cite{atwater2010plasmonics}, and thermoelectrics\,\cite{hochbaum2008enhanced,boukai2008silicon}.

\section{Methodology}

The frequency shift due to phonon scattering is given by the real part of the self-energy \cite{Maradudin1962pr,Lang1999,PRB2003MgB2,Bonini2007,Turney2009prb1,Turney2009thesis}. In the first order, the only contribution is the bubble diagram \cite{Maradudin1962pr,PRB2003MgB2}, which corresponds to the four-phonon scattering term $\Delta\omega_\lambda^{(4)}$ in Eq.(\ref{Delta4}). Here $\lambda$ is short for $(\mathbf{q},\nu)$ with $\mathbf{q}$ and $\nu$ representing phonon wave vector and dispersion branch, respectively. In the second order, the only contribution is the loop diagram \cite{Maradudin1962pr,PRB2003MgB2}, which is the three-phonon scattering term $\Delta\omega_\lambda^{(3)}$ in Eq.(\ref{Delta3}). In some literature \cite{Lang1999,Turney2009prb1,Turney2009thesis}, a tadpole diagram was included, which is the term of the three-phonon process to the second order $\Delta\omega_\lambda^{(T)}$. This term is generally negligible. For instance, Lazzeri \textit{et al.}\cite{PRB2003MgB2} pointed out that this term is zero in their system due to the lattice translational invariance. The expressions of $\Delta\omega_\lambda^{(3)}$, $\Delta\omega_\lambda^{(4)}$ and $\Delta\omega_\lambda^{(T)}$ are given by
\begin{widetext}
\begin{gather}
\Delta\omega_\lambda^{(4)} = \frac{\hbar}{8N_{\mathbf{q}}\omega_\lambda} \sum_{\lambda_1} V_{-\lambda\lambda\lambda_1-\lambda_1}^{(4)}\cdot\frac{2n_1+1}{\omega_1}, \label{Delta4}\\
\Delta\omega_\lambda^{(3)} = \frac{\hbar}{16N_{\mathbf{q}}\omega_\lambda} \sum_{\lambda_1\lambda_2} \left[\left|V_{\lambda\!-\!\lambda_1\!-\!\lambda_2}^{(3)}\right|^2 \! \delta_{\mathbf{q}\!-\!\mathbf{q}_1\!-\!\mathbf{q}_2} \! \cdot \!\frac{n_1+n_2+1}{\omega_1\omega_2(\omega_\lambda\!-\!\omega_1\!-\!\omega_2)_P} + 2\left|V_{\lambda\lambda_1\!-\!\lambda_2}^{(3)}\right|^2 \! \delta_{\mathbf{q}\!+\!\mathbf{q}_1\!-\!\mathbf{q}_2} \!\cdot \!\frac{n_1-n_2}{\omega_1\omega_2(\omega_\lambda\!+\!\omega_1\!-\!\omega_2)_P}\right], \label{Delta3}\\
\Delta\omega_\lambda^{(T)} = \frac{\hbar}{8N_{\mathbf{q}}\omega_\lambda} \sum_{\lambda_1}\sum_{\nu_2} V_{\lambda,-\lambda,(\mathbf{0},\nu_2)}^{(3)}V_{\lambda_1,-\lambda_1,(\mathbf{0},\nu_2)}^{(3)}\cdot\frac{2n_1+1}{\omega_1\omega_2(\omega_2)_P}, \label{DeltaT}
\end{gather}
where $V_{\pm}^{(3)}$ and $V_{\pm\pm}^{(4)}$ are the three-phonon and four-phonon scattering matrices given by 
\begin{gather}
V_{\lambda\lambda_1\lambda_2}^{(3)}\!=\!\sum_{b\!,l_1b_1\!,l_2b_2}\sum_{\alpha\alpha_1\!\alpha_2}\Phi_{0b\!,l_1b_1\!,l_2b_2}^{\alpha\alpha_1\alpha_2}\frac{e_{\alpha b}^\lambda e_{\alpha_1 b_1}^{\lambda_1} e_{\alpha_2 b_2}^{\lambda_2}}{\sqrt{m_b m_{b_1} m_{b_2}}} \psi_3, \label{V3}\\
V_{\lambda\lambda_1\lambda_2\lambda_3}^{(4)}=\sum_{b,l_1b_1,l_2b_2,l_3b_3}\sum_{\alpha\alpha_1\alpha_2\alpha_3}\Phi_{0b,l_1b_1,l_2b_2,l_3b_3}^{\alpha\alpha_1\alpha_2\alpha_3}\frac{e_{\alpha b}^\lambda e_{\alpha_1 b_1}^{\lambda_1} e_{\alpha_2 b_2}^{\lambda_2} e_{\alpha_3 b_3}^{\lambda_3}}{\sqrt{m_b m_{b_1} m_{b_2} m_{b_3}}} \psi_4 . \label{V4}
\end{gather}
\end{widetext}

It should be noted that, in phonon frequency shifts, the four-phonon term makes the lowest-order (or main) contribution, which is generally much larger than the three-phonon term. This is different from phonon scattering rates, in which the four-phonon term is in a higher order, which makes generally much smaller contribution than the three-phonon term\cite{Feng_fourphonon1,Feng_BAs}.

In the equations above, $n=(e^{\hbar\omega/k_BT}-1)^{-1}$ is the phonon occupation number, $\omega$ is the phonon frequency, and $e$ is the phonon eigenvector. $l$, $b$, and $\alpha$ label the indices of primitive cells, basis atoms, and the ($x$,$y$,$z$) directions, respectively. $\mathbf{r}_l$ is the position of the primitive cell $l$. The Cauchy principle value $1/(x)_P=x/(x^2+\epsilon^2)$, where the infinitesimal $\epsilon$ is a broadening factor.

The phases $\psi_3$ and $\psi_4$ should keep consistent with the phase used in the dynamical matrix $D$ that solves eigenvectors. Either of the following two sets of equations can be used:

\begin{subequations}
	\label{Eq_ratom}
	\begin{align}
	\psi_3 &= e^{i (\mathbf{q}\cdot \mathbf{r}_{0b} +  \mathbf{q}_1\cdot \mathbf{r}_{l_1b_1} +  \mathbf{q}_2\cdot \mathbf{r}_{l_2b_2})}, \label{Eq_psi3_ratom} \\
	\psi_4 &= e^{i (\mathbf{q}\cdot \mathbf{r}_{0b} +  \mathbf{q}_1\cdot \mathbf{r}_{l_1b_1} +   \mathbf{q}_2\cdot \mathbf{r}_{l_2b_2} +  \mathbf{q}_3 \cdot \mathbf{r}_{l_3b_3})},  \label{Eq_psi4_ratom} \\
	D_{\alpha\alpha_1}^{bb_1}(\mathbf{q}) &=\frac{1}{\sqrt{m_bm_{b_1}}}\sum_{l_1}\Phi_{0b,l_1b_1}^{\alpha\alpha_1} e^{i\mathbf{q}\cdot(\mathbf{r}_{l_1b_1}-\mathbf{r}_{0b})} ; \label{Eq_dyn_ratom}
	\end{align}
\end{subequations}

\begin{subequations}
	\label{Eq_rcell}
	\begin{align}
	\psi_3 &= e^{ i (\mathbf{q}_1\cdot \mathbf{r}_{l_1} +  \mathbf{q}_2\cdot \mathbf{r}_{l_2}) }, \label{Eq_psi3_rcell}\\
	\psi_4 &= e^{ i (\mathbf{q}_1\cdot \mathbf{r}_{l_1} +   \mathbf{q}_2\cdot \mathbf{r}_{l_2} +  \mathbf{q}_3 \cdot \mathbf{r}_{l_3}) }, \label{Eq_psi4_rcell}\\
	D_{\alpha\alpha_1}^{bb_1}(\mathbf{q}) &=\frac{1}{\sqrt{m_bm_{b_1}}}\sum_{l_1}\Phi_{0b,l_1b_1}^{\alpha\alpha_1} e^{i\mathbf{q}\cdot\mathbf{r}_{l_1}}. \label{Eq_dyn_rcell}
	\end{align}
\end{subequations}

The first set of phases (Eq.(\ref{Eq_ratom})) uses the positions of atoms $r_{lb}$, while the second (Eq.(\ref{Eq_rcell})) uses the positions of cells $r_{l}$. We have verified numerically that the two sets of equations give the same results. However, it should be noted that the mixing of the two sets may give wrong results. For example, if the eigenvectors are calculated by using the dynamical matrix in Eq.(\ref{Eq_dyn_ratom}) (as implemented in Phonopy\cite{Phonopy}), the phases $\psi_3$ and $\psi_4$ calculated by $r_l$ (Eqs.(\ref{Eq_psi3_rcell}) and (\ref{Eq_psi4_rcell})) will give wrong three and four-phonon scattering rates as well as frequency shifts (see the Appendix \ref{appen_phase}.)

The second-, third- and fourth-order force constants $\Phi$ were obtained by first principles. In Eq.(\ref{Delta4}), although $V^{(4)}$ is a complex number, we have verified that the summation $\sum_{\lambda_1}$ cancels out the imaginary part and give a real $\Delta\omega^{(4)}_\lambda$.

%



It is noted that the above methods do not include the effect of thermal expansion. As pointed out by Bonini \textit{et al.}\cite{Bonini2007}, the thermal expansion contribution should be taken into account to obtain the total frequency shift:
\begin{equation}
\Delta\omega_\lambda^{\rm tot}=\Delta\omega_\lambda^{(4)}+\Delta\omega_\lambda^{(3)}+\Delta\omega_\lambda^{(T)}+\Delta\omega_\lambda^{\rm quasi}. \label{eq1}
\end{equation}

\begin{figure}[b]
	\centering
	\includegraphics[width= 3.3in]{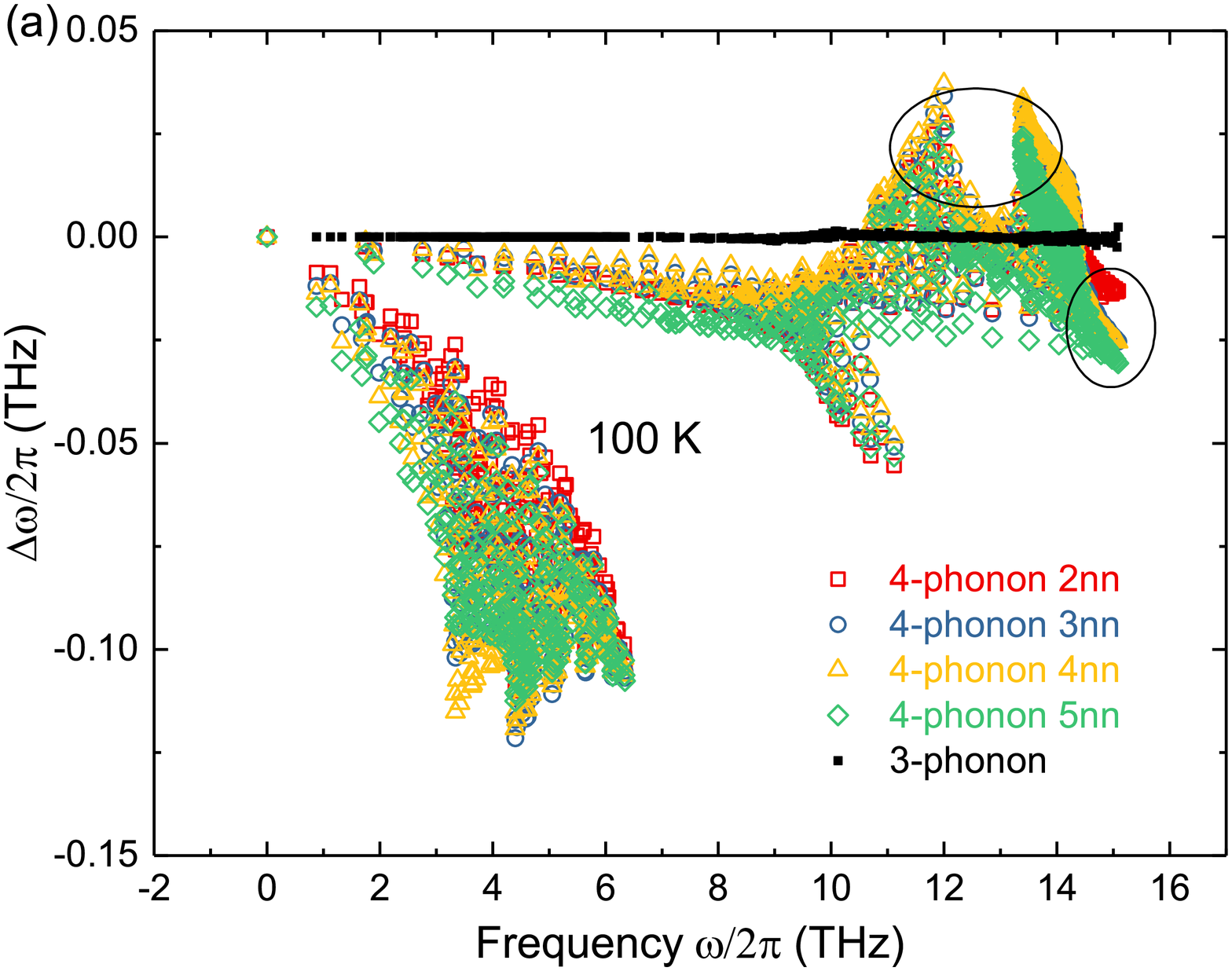}
	\includegraphics[width= 3.3in]{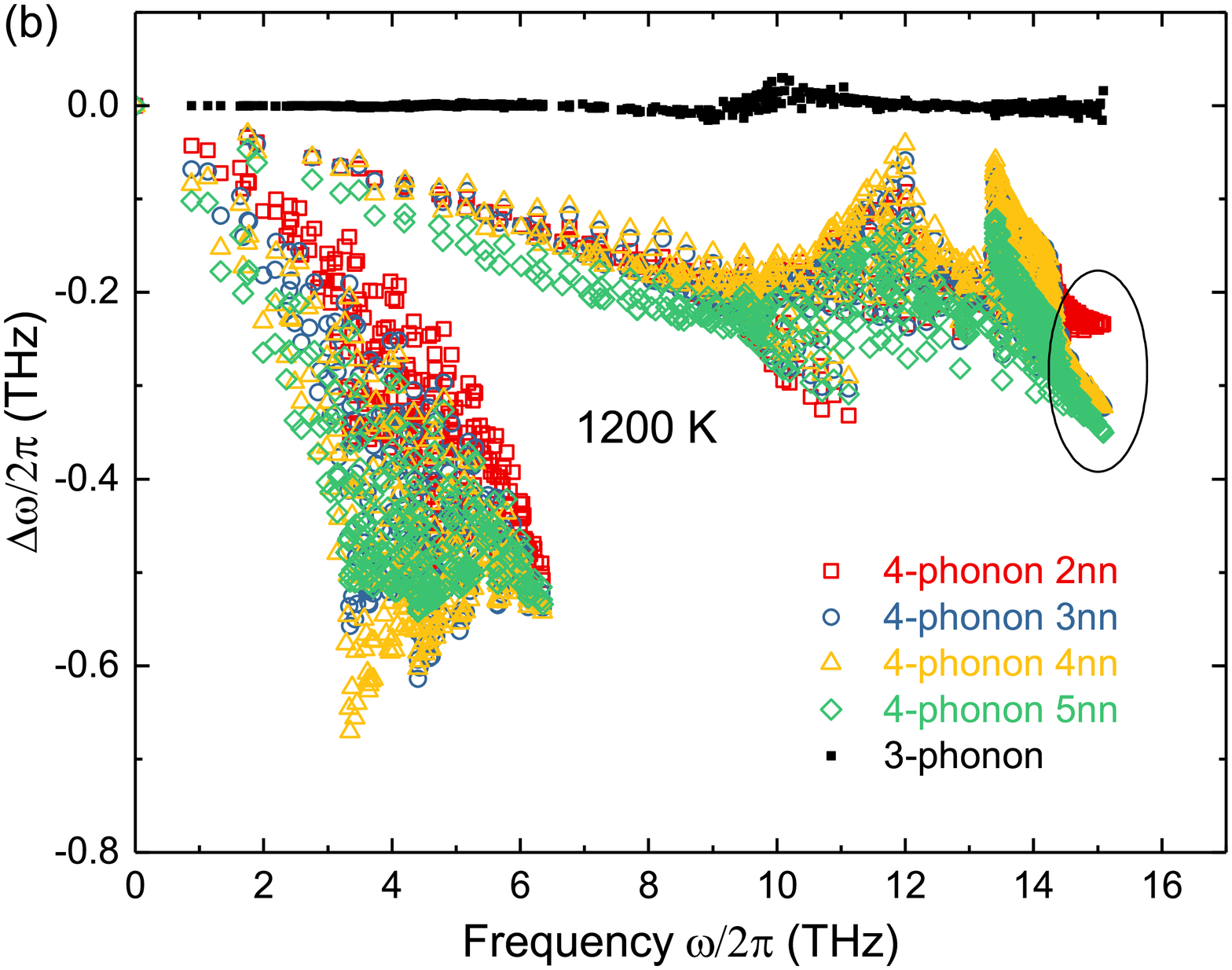}
	\caption{Phonon frequency shifts of Si at 100 K (a) and 1200 K (b) by four-phonon (colored open dots) and three-phonon scattering (black solid dots). The effect of fourth-order force constant cutoff radius is studied, as labeled by the nearest neighbors (nn).}\label{fig_nn}
\end{figure}

\begin{figure*}[t]
	\centering
	\includegraphics[width= 6.3in]{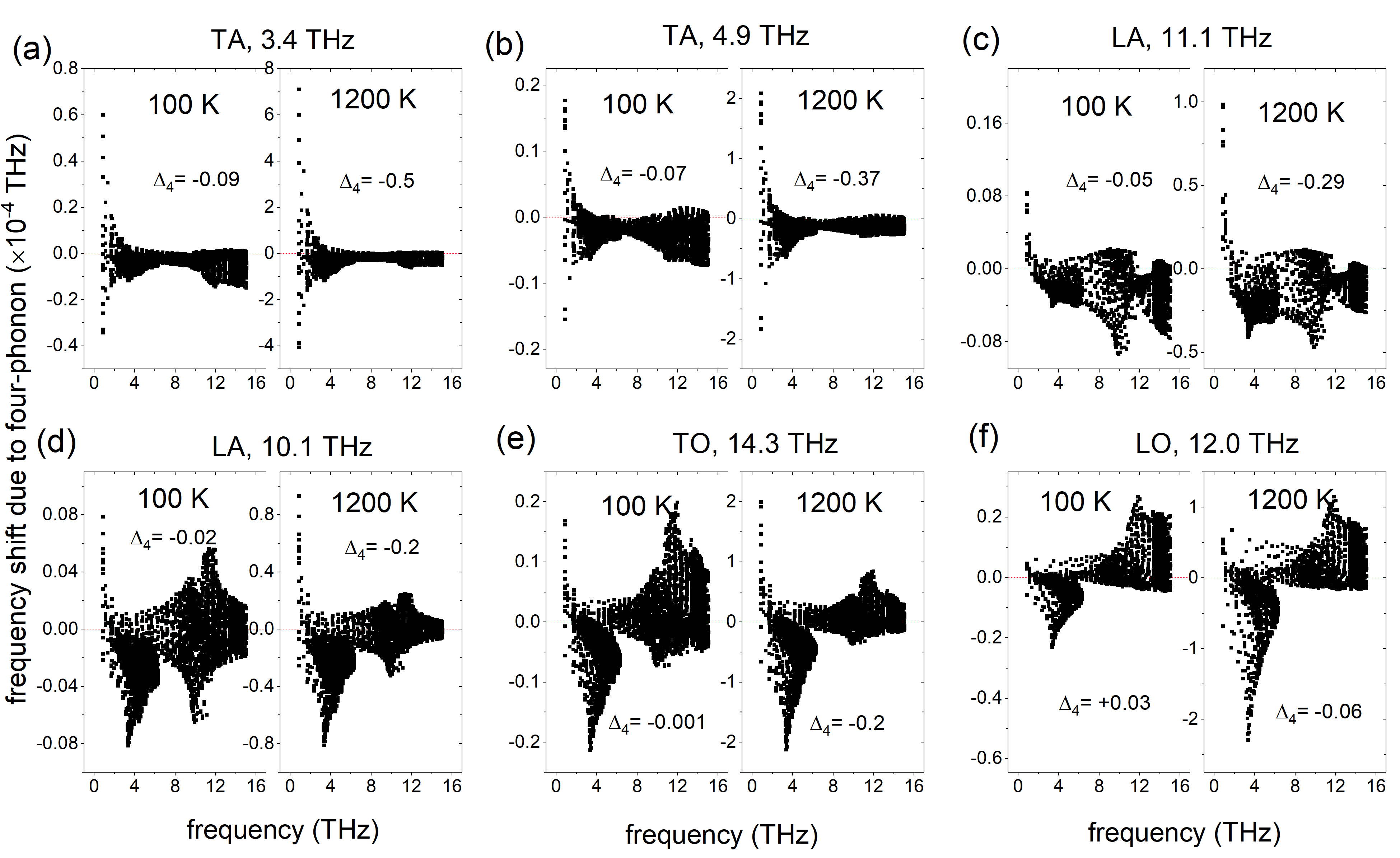}
	\caption{Four-phonon frequency shift phase spaces of six phonon modes examples. Each panel contains two temperatures, 100 K and 1200 K. The phonon branch and frequency is labeled on the top of each panel. The total four-phonon shift of each mode is the summation of the shifts induced by all the other phonon modes. Panels (a)-(e) show negative overall shift at low and high temperatures. Panel (f) shows positive overall shift at low temperature and negative shift at high temperature.}\label{phase_space}
\end{figure*}

\begin{figure*}[t]
	\centering
	\includegraphics[width= 6.6in]{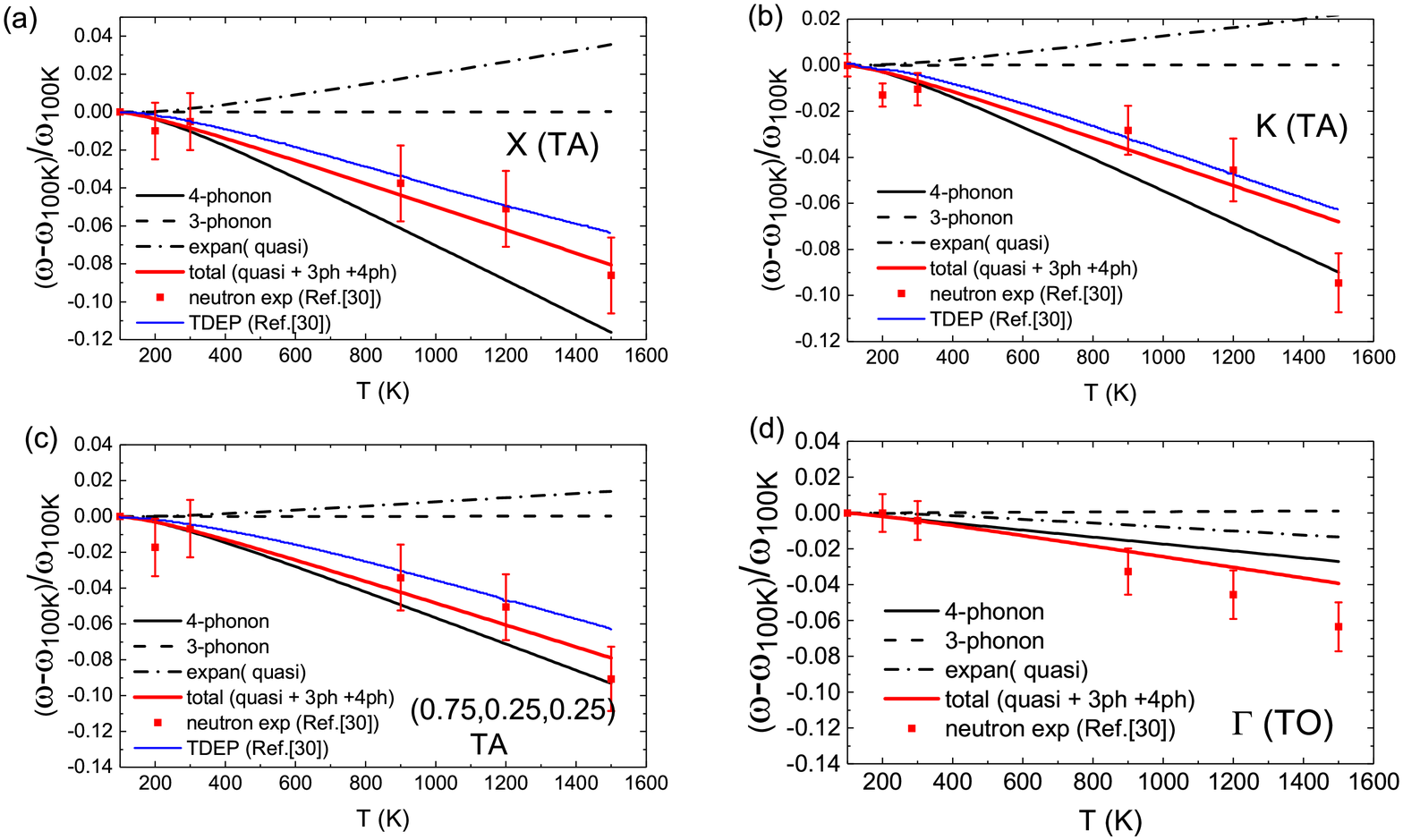}
	\caption{The frequency shifts calculated from first principles compared to the experimental inelastic neutron scattering data. The modes for demonstration are the lower-branch TA phonons at the high-symmetry $\mathbf{q}$ points: $X$ (a) and $K$ (b). Another point at $2\pi/a\cdot$(0.75,0.25,0.25) located away from high-symmetry lines is shown as a reference in (c). The optical phonon at the $\Gamma$ point is shown in (d). The frequency shift due to thermal expansion (dash-dotted line), 4-phonon scattering (dotted line), 3-phonon scattering (dashed line) and the total effects (solid red line) are calculated by this work. The experimental data (red dot) as well as the temperature-dependent effective potential (TDEP) calculation results (solid blue line) given by Ref.\cite{Kim2018} are shown for comparison. No TDEP data are available for the $\Gamma$ point.}\label{fig_TA}
\end{figure*}
The second-order force constant and phonon frequencies were calculated by Phonopy\cite{Phonopy} using density functional perturbation theory (DFPT) based on VASP\cite{VASP}. The supercell is taken as $4\times4\times4$ primitive cells. The third-order force constants were calculated via Thirdorder, a package of ShengBTE, based on VASP using $4\times4\times4$ primitive cells considering up to the sixth nearest neighbor. The fourth-order force constant considering up to the second and third nearest neighbors were calculated using $4\times4\times4$ primitive cells, while those considering up to the fourth and fifth nearest neighbors were calculated using $5\times5\times5$ primitive cells.  We used the Perdew-Burke-Ernzerhof (PBE) parameterization of the generalized gradient approximation (GGA) for exchange and correlation functionals \cite{PBE1,PBE2} with projector-augmented-wave method\cite{PAW1}. The plane-wave energy cutoff is 319 eV, and the electron $\mathbf{k}$ mesh is taken as 4$\times$4$\times$4. The energy convergence threshold is set at $10^{-7}$ eV. The phonons are taken with a $16\times16\times16$ $\mathbf{q}$-mesh in the BZ. The calculation of quasiharmonic frequency requires the temperature-dependent lattice constant or the thermal expansion coefficient, which can be calculated by the volume derivative of phonon entropy. The details of the calculations of thermal expansion coefficient and quasi-harmonic frequency shift are shown in Appendix \ref{appen_expansion}.



\section{Results and Discussions}

The phonon frequency shifts $\Delta\omega_\lambda^{(3)}$ and $\Delta\omega_\lambda^{(4)}$ as a function of frequency are shown in Fig.\ref{fig_nn}. Since the $\Delta\omega^{(T)}$ term is found to be zero, it is not shown in the figure. It is seen that the four-phonon scattering overwhelmingly dominates the frequency shifts. As pointed out by Marududin and Fein\,\cite{Maradudin1962pr}, the four-phonon scattering softens the acoustic phonon modes. Interestingly, we find that the softening of transverse acoustic (TA) mode is significantly larger than the others, although the TA phonons have much longer relaxation time. It is probably because that the TA mode is the softest as seen in Fig.\ref{fig_gru} that the Gr\"{u}neisen parameter of TA mode is largest. At room temperature, the relative shifts of the longitudinal acoustic (LA) and all the optical modes are around 0.4\%, while those for the TA mode are 0.5-2.5\%. As the temperature increases to 1200 K, LA and optical branches shift down by $\sim$2.8\%, while the TA branch by 2-6.5\%. Based on this observation, we conclude that the phonon scattering rate (imaginary part of self-energy) and the frequency shift (real part of self-energy) are not necessarily positively correlated.

Interestingly, we note that the phonon anharmoncity does not always soften phonon modes. It is well known from the anharmonic potential well that anharmonicity deviates the spring constant from the harmonic parabolic potential well to a softer mode. However, we find that at low temperature, some optical phonon modes near the BZ boundary can be stiffened by anharmonicity, which was not considered in Ref.\,\cite{Maradudin1962pr}. To gain physical insight into this phenomenon, we have looked into the four-phonon scattering phase space. Some examples are shown in the Fig.\ref{phase_space}. The results are based on Eq.(\ref{Delta4}). Each panel shows a phonon mode, and it contains the frequency shifts induced by all the other modes. For example, for the first TA mode at 3.4 THz shown in Fig.\ref{phase_space} (a), although some other modes bring its frequency up, most other modes bring its frequency down. As a result, the total four-phonon frequency shift of this TA mode is negative (-0.09 THz at 100 K, and -0.5 THz at 1200 K). Similar trends are found in Figs.\ref{phase_space} (b)-(e). For the LO mode at 12 THz in Fig.\ref{phase_space} (f), the low-frequency phonons bring its frequency down while high-frequency phonons bring its frequency up. At low temperature, the overall four-phonon frequency shift is positive (+0.03 THz at 100 K). As temperature increases, the low-frequency phonons' impact increases faster than the high-frequency phonons' impact. As seen at 1200 K, the low-frequency phonons' negative shifts dominate over high-frequency phonons' positive shifts, and the overall shift becomes negative (-0.06 THz at 1200 K). The temperature effect is reflected in the factor $(2n+1)/\omega$ in Eq.\ref{Delta4}. It is easy to find that the function  $\frac{1}{\omega} \left( \frac{2}{\exp (\hbar\omega/k_BT)-1} +1 \right)$ increases with $T$, and low $\omega$ increases faster than high $\omega$ with increasing $T$.

Furthermore, the effect of the cutoff radius of the fourth-order force constants is studied. We find that increasing the cutoff radius from the second to the third nearest neighbor can substantially increase the frequency shift, especially for the high frequency phonons. This phenomenon is quite distinct from the four-phonon scattering rates, which are not sensitive to the fourth-order force constants cutoff radius, especially for the high-frequency phonons\cite{Feng_BAs}.


The phonon frequency shifts as a function of temperature at four given BZ points are shown in Fig.\ref{fig_TA}. The four panels correspond to the TA modes at the $X$, $K$, and (0.75,0.25,0.25) points and the TO mode at $\Gamma$ point in the Brillouin zone, respectively. At low temperature, the effects of thermal expansion and three-phonon scattering are negligible, and the total frequency shift comes from the four-phonon scattering which is the first order perturbation. At a higher temperature, the thermal expansion effect becomes important, though not as important as the four-phonon scattering, while the three-phonon effect remains negligible since it is a higher-order term. Our results are compared to the existing data from first principles simulations and inelastic neutron scattering experiment\,\cite{Kim2018}. Due to the negative Gr\"{u}neisen parameter of the TA branch and the positive thermal expansion of silicon at room to high temperatures, $\Delta\omega^{\rm quasi}$ is positive while the others are negative. Our results show that the anharmonic phonon shift increases linearly with temperature at high temperature, which agrees well with the statement predicted by Marududin and Fein\,\cite{Maradudin1962pr}. 
Our prediction agrees reasonably well with the experimental data. 

\begin{figure}[t]
	\centering
	\includegraphics[width= 3.3in]{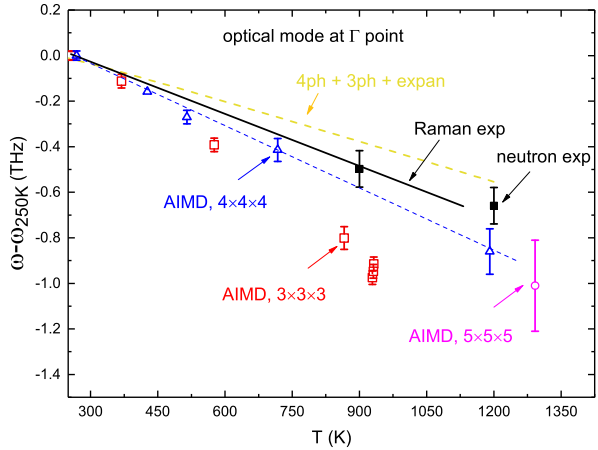}
	\caption{The phonon frequency shift of the optical phonon at the $\Gamma$ point as a function of temperature. The normal mode analysis predictions based on AIMD with are compared to the Raman\cite{Si_Raman1,Si_Raman2} and neutron\cite{Kim2018} experimental data.}\label{fig_AIMD}
\end{figure}

The perturbation theory has a limitation since it does not include higher order anharmonicities, which may account for the discrepancy between the prediction of experiment for the optical phonon at $\Gamma$ point.
To understand this discrepancy, we have conducted a separate  normal mode analysis\cite{Feng2014Jn,Feng2015jap,Koker2009prl,Thomas2010prb} calculation based on ab initio molecular dynamics (AIMD). To ensure the AIMD simulations correctly capture the thermal expansion contribution, the lattice constants are adjusted at each temperature to ensure the pressure of the micro canonical (NVE) ensemble is zero. The time step in the simulation is set to 1.5 fs, which is short enough to resolve all the phonon modes in Si. Before each NVE simulation, canonical-ensemble (NVT) simulations were conducted to equilibrate the system. The total NVE simulation time at each temperature is run for 40,000 steps with the first 5,000 steps discarded as the initialization steps and the remaining 35,000 steps taken as the MD trajectory for the spectral energy density (SED) analysis. The obtained SED for each phonon mode is a distinct peak (see Appendix \ref{appen_SED}), from which the anharmonic phonon frequency can be extracted. 

Here we use NVT instead of NPT before NVE simulations. NPT followed by NVE is usually used in classical MD simulations for normal mode analysis. But in AIMD, the simulation domain is small and the MD fluctuation ($\sim1/\sqrt{N}$) is large. In the pressure vs. time plot in NPT simulations, we find that the pressure oscillates significantly around zero although the average is zero. And when we switch the NPT to NVE at some time point, the pressure at that time point is typically large non-zero because it is oscillating. As a result, the following NVE simulation has an average large non-zero pressure. (Normal mode analysis requires zero-pressure to eliminate the pressure effect on the phonon properties.) Therefore, we have to spend much more effort to use NVT instead of NPT before NVE: we manually adjust the lattice constants and run NVT for them, and then find out which lattice constant gives zero-pressure for a given temperature. Then, we use that lattice constant to run NVE. Only by this way can we guarantee the pressure in the NVE simulation is zero. Note that due to the large fluctuation, the lattice constant read out from NPT simulations has a large error bar, and it is not accurate enough to give zero pressure in NVE simulations. This is why the AIMD simulations for normal mode analysis are time expensive.

In Fig.\ref{fig_AIMD}, the predictions with $3\times3\times3$ and $4\times4\times4$ simulation cells are compared to the Raman\cite{Si_Raman1,Si_Raman2} and neutron\cite{Kim2018} experimental data. We find that the size effect of AIMD simulations from $3\times3\times3$ to $4\times4\times4$ is strong, probably because we used the $\Gamma$ point only as the electronic $\mathbf{k}$-mesh in the AIMD simulations. For $3\times3\times3$ supercell, the $\Gamma$ point only might not be able to predict interatomic force accurately. The larger the domain size is, the better the $\Gamma$-only $\mathbf{k}$-mesh works in DFT. It can be seen that the size effect from $4\times4\times4$ to $5\times5\times5$ is smaller. The AIMD simulation results agree reasonably well with experiment with a slight overestimation, which is suspected to be originated from the fast algorithm of AIMD simulations that might sacrifice the accuracy of DFT electronic method. Since AIMD simulations naturally include all the orders of anharmonicity, the underestimation of frequency shift by four-phonon scattering might originate from the ignorance of higher order scattering, which is beyond the scope of our study. 

Phonon softening can affect the thermal conductivity of materials, which is important for thermoelectric applications. Thermoelectric materials such as SnSe\cite{PhysRevB.94.125203}, Bi$_2$Te$_3$\cite{Wang2013jht}, and PbTe\cite{PbTe_Xia} typically have significant phonon softening effects at high temperatures due to their long-range Coulomb and resonant bonds. SnSe even has a phase change at high temperature. Wang \textit{et al.} have shown that the Bi$_2$Te$_3$ could have a large phonon frequency shift even at room temperature by using classical MD simulations with SED analysis \cite{Wang2013jht}. The phonon softening affects the thermal conductivity by changing the phonon group velocity, phonon-phonon scattering rates and phonon specific heat. Typically the most straight-forward and significant impact is on the group velocity. Recently it was found that the phonon-phonon scattering rate could also be substantially affected in some system\cite{PbTe_Xia}. 



\section{Conclusions}
In summary, we have calculated the temperature-dependent phonon frequencies of silicon in the full Brillouin zone by first principles perturbation theory. The frequency shift as compared to 0 K includes the effects of  anharmonic phonon scattering and thermal expansion, with the former found to be dominating. The total frequency shifts increase linearly with temperature at high temperature. At 1200 K, the TA branch shifts down by 2-6.5\%, and the other branches shift down by about 2.8\%. Our results are found to agree well with the existing inelastic neutron scattering data. 
The calculation of fourth order force constant as well as the frequency shifts are not expensive, and we expect our method and results will inspire the calculations of other materials such as the thermoelectric materials in which the phonon softening effect is strong.

\begin{acknowledgements}
Simulations were preformed at the Rosen Center for Advanced Computing of Purdue University. The work was supported by the Defense Advanced Research Projects Agency (Award No. HR0011-15-2-0037). 
\end{acknowledgements}

\appendix

\section{Phase issue in determining the four-phonon frequency shift}
\label{appen_phase}

\begin{figure}[tbph]
	\centering
	\includegraphics[width= 3.4in]{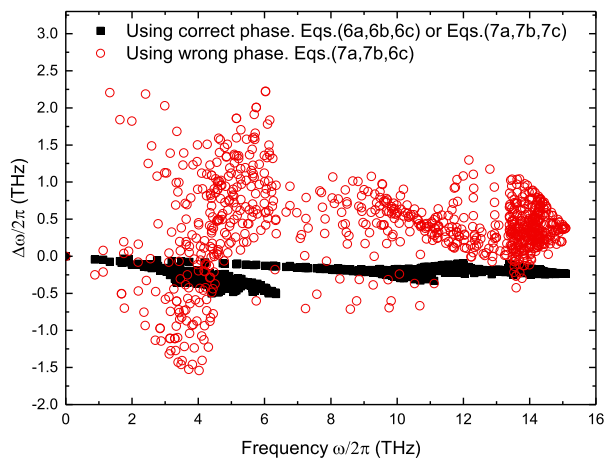}
	\caption{Phonon frequency shifts at 1200K calculated from correct and wrong phases.}\label{fig_rcell}
\end{figure}

\section{Thermal expansion contribution to phonon shifts}
\label{appen_expansion}

The calculation of quasiharmonic frequency requires the temperature-dependent lattice constant or the thermal expansion coefficient, which can be calculated by the volume derivative of phonon entropy. With the detailed derivation process explained in Ref.\,\cite{Argaman2016}, here we only show the neat procedure. Starting from DFT, the harmonic phonon frequencies $\omega_\lambda^0$ and the pressures at different lattice constants around $a_0$ are calculated. $a_0$ is the lattice constant at 0 K. From the phonon frequency response to the volume change, the mode-dependent Gr\"{u}neisen parameter $\gamma_\lambda$ is defined as 
\begin{equation}
\label{eq_Gru}
\gamma_\lambda=-\frac{V}{\omega_\lambda^0}\frac{\partial \omega_\lambda^0}{\partial V}.
\end{equation}
Gr\"{u}neisen parameter is calculated by using finite difference of the phonon frequencies at several different volumes. Since changing the volume is equivalent to changing temperature, the Gr\"{u}neisen parameter is therefore ``temperature-averaged''. Nevertheless, we find that the Gr\"{u}neisen parameter is almost identical at different volumes, which indicates that it is temperature-independent for Si. From the pressure response to the volume change, the bulk modulus $B$ can be calculated as
\begin{equation}
\label{eq_modulus}
B=-V\frac{dP}{dV}.
\end{equation}
The obtained $\gamma_\lambda$ and $B$ are then used as the inputs of the calculation of the temperature-dependent thermal expansion coefficient
\begin{equation}
\label{eq_alpha}
\alpha_V(T)=-\frac{k_B}{N_\mathbf{k}V_{\rm cell}B}\sum_{\lambda}\gamma_\lambda\cdot\left(\frac{x}{2}\right)^2\cdot\left(1-\coth^2(\frac{x}{2})\right)
\end{equation}
and the temperature dependent volume
\begin{equation}
\label{eq_V}
V(T)=V(0)\cdot\exp\left(\int_0^T \alpha_V(T)dT\right).
\end{equation}
Here $N_\mathbf{q}$ is the number of $\mathbf{q}$ points. $V_{\rm cell}$ is the volume of a primitive cell, which is $a^3/4$ for silicon. $k_B$ is the Boltzmann constant. $x$ is short for $\hbar\omega/k_BT$. Finally, the obtained $\alpha_V$ and $\gamma_\lambda$ are used as the input to calculate the temperature-dependent quasiharmonic frequency 
\begin{equation}
\label{eq_quasi}
\omega_\lambda^{\rm quasi}(T)=\omega_\lambda^0\!\cdot\!\left(\frac{V(T)}{V(0)}\right)^{-\gamma_\lambda}=\omega_\lambda^0\!\cdot\!\left[\exp\left(\int_0^T\! \alpha_V(T)dT\right)\right]^{-\gamma_\lambda}.
\end{equation}
Some literature used a different approach, by minimizing the Helmholtz free energies at give temperatures\,\cite{Mounet2005,Bonini2007,Souvatzis2007}, to calculate the temperature-dependent lattice constant. The two approaches, in principle, give the same results.

\begin{figure}[t]
	\centering
	\includegraphics[width= 3.4in]{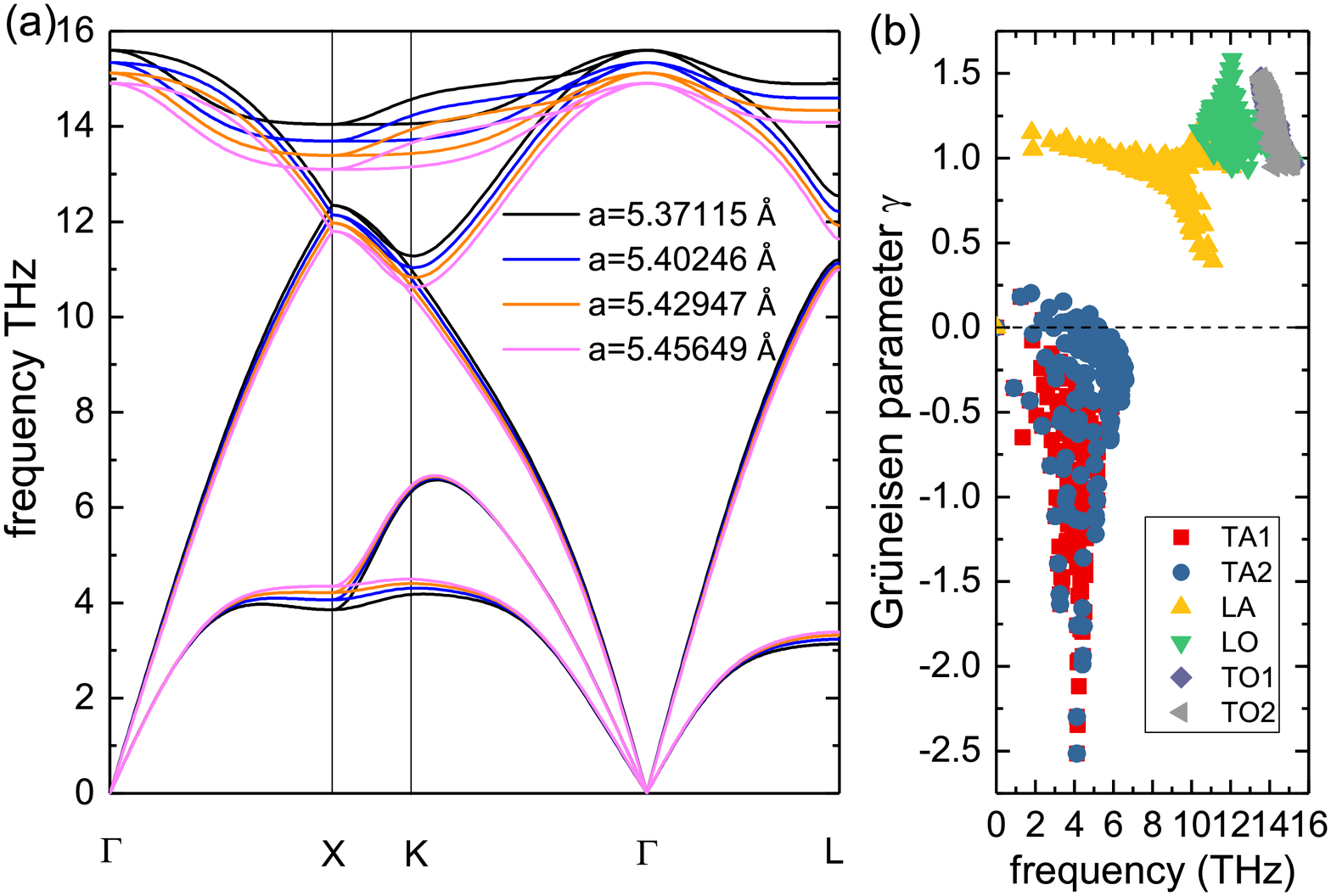}
	\caption{Si phonon dispersion relations at different lattice constants and the mode-resolved Gr\"{u}neisen parameter calculated from first principles.}\label{fig_gru}
\end{figure}

\begin{figure}[t]
	\centering
	\includegraphics[width= 3.4in]{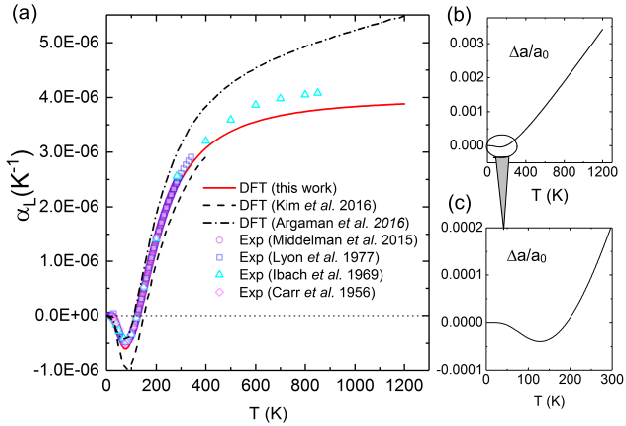}
	\caption{(a) Linear thermal expansion coefficients calculated from quasi-harmonic approximation by this work (solid red line), Kim \textit{et al.}\cite{Kim2018}, and Argaman \textit{et al.}\cite{Argaman2016}, with comparison to the experiment by Middelman \textit{et al.}\cite{Middelmann2015}, Lyon \textit{et al.}\cite{Lyon1977}, Ibach \textit{et al.}\cite{Ibach1969}, and Carr \textit{et al.}\cite{Carr1956}. The deduced lattice change $\Delta a/a_0$ is plotted in (b), with the low-temperature data enlarged in (c).}\label{fig_exp}
\end{figure}

The harmonic frequencies of a $16\times16\times16$ $\mathbf{q}$-mesh are obtained as a function of lattice constant, with the dispersion along high-symmetry directions shown in Fig\,\ref{fig_gru} (a). It is seen that the longitudinal acoustic (LA) and all the optical phonon modes are softened as the lattice constant increases. Their Gr\"{u}neisen parameters are thus positive, as shown in Fig\,\ref{fig_gru} (b). In contrast, most phonons of the transverse acoustic (TA) branch are stiffened as the lattice constant increases, and thus their Gr\"{u}neisen parameters are negative. This phenomenon has been discussed extensively in the literature\cite{Liu2011,Fang2014,Dove2016,Argaman2016,Sun2016}. The bulk modulus $B$ of Si calculated in this work is 97.61 GPa, which matches well with the experimental value 97.8 GPa \cite{hopcroft2010young}. Based on the Gr\"{u}neisen parameters and the bulk modulus, we obtained the linear thermal expansion coefficient $\alpha_L$ as a function of temperature as shown in Fig.\,\ref{fig_exp} (a). For isotropic materials, $\alpha_L=\alpha_V/3$. Resulting from the negative Gr\"{u}neisen parameter of the TA mode, silicon exhibits negative thermal expansion coefficient at low temperature. As the temperature increases, more LA and optical phonons are excited, and the overall thermal expansion coefficient turns to positive. The transition temperature is predicted at about 128 K, which agrees well with the experimental value at around 125 K\,\cite{Middelmann2015,Lyon1977,Ibach1969,Carr1956}. As the temperature increases far beyond the Debye temperature, all the modes tend to be excited, and $\alpha_V$ tends to saturate. As compared to the first principles prediction in literature\,\cite{Argaman2016,Kim2018}, our result agrees much better with experiment\,\cite{Middelmann2015,Lyon1977,Ibach1969,Carr1956}. In Figs.\,\ref{fig_exp} (b,c), the relative lattice constant change is plotted as a function of temperature. The lattice constant we obtained at 0 K is 5.4037 \AA. At 128 K, the lattice constant shrinks the most, by 0.004\%, which is a subtle change. At room temperature and 1200 K, the lattice constant is expanded by 0.021\% and 0.341\%, respectively. These results agree well with experiment\,\cite{yim1974thermal,okada1984precise}.

\section{Example of spectral energy density from AIMD}
\label{appen_SED}

\begin{figure}[tbph]
	\centering
	\includegraphics[width= 3.4in]{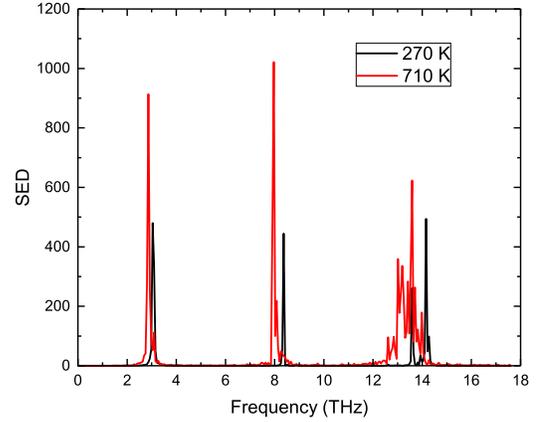}
	\caption{Spectral energy density of Si extracted from AIMD simulations at two different temperatures.}\label{fig_SED}
\end{figure}



\clearpage

%


\end{document}